\documentclass[aps,pra,twocolumn]{revtex4-2}

\usepackage{color,natbib,bm}
\newcommand{\tcr}[1]{\textcolor{black}{#1}}

\usepackage{graphicx,amsmath,amssymb}

\begin{document}

\title{Maximum heralding probabilities of nonclassical-state generation from a two-mode Gaussian state via photon-counting measurements}

\author{Jaromír Fiurášek}
\affiliation{Department of Optics, Faculty of Science, Palack\'y University, 17.\ listopadu 12, 77900  Olomouc, Czech Republic}

\begin{abstract}
Highly nonclassical states of light --  such as the approximate Gottesman-Kitaev-Preskill states, states exhibiting cubic nonlinear squeezing,  or cat-like states -- can be generated from experimentally accessible Gaussian states via photon counting measurements on selected modes, conditioned on specific outcomes of these heralding events. A simplest yet important example of this approach  involves performing photon number measurements on one mode of a two-mode entangled Gaussian state. The heralding probability of this scheme is a key figure of merit, as it determines the generation rate of the target nonclassical state. In this work we show that the maximum heralding probability for the two-mode setting  can be calculated analytically, and we investigate its dependence on the number of detected photons $n$. Our results show that the number of required experimental trials scales only polynomially with  $n$. Generation of highly complex optical quantum states with high stellar rank is thus in principle possible in this setting, given access to sufficiently strong squeezing.

\end{abstract}

\maketitle

\section{Introduction}

Gaussian boson sampling has recently attracted considerable attention \cite{Lund2014,Hamilton2017,Paesani2019,Zhong2020,Zhong2021,Thekkadath2022,Oh2024}. Besides representing a specific limited quantum computing model, it also offers a promising and feasible route to generation of highly nonclassical quantum states of light for applications in optical quantum technologies \cite{Su2019,Takase2023,Larsen2025,Hanamura2025,Takase2026}. In conditional state preparation via Gaussian boson sampling a multimode entangled Gaussian quantum state is prepared and some of the modes are measured in Fock basis. Detection of specific numbers of photons heralds preparation of a
target state in the unmeasured modes. This framework in fact encompasses a wide range of experimental setups \cite{Lvovsky2020,Biagi2022} including quantum-state engineering schemes relying on conditional addition \cite{Zavatta2004,Barbieri2010,Kumar2013,Fadrny2024,Chen2024} or subtraction \cite{Ourjoumtsev2006,Nielsen2006,Wakui2007,Nielsen2010,Huang2015,Endo2025} of photons.

 In the past, the experimental schemes were mainly  designed to be robust with respect to inefficient detection, which was typically achieved at the expense of reduced success probability. 
 However, the development of highly efficient superconducting \tcr{detectors achieving photon number resolution} \cite{Lita2008,Fukuda2011,Eaton2023,Cheng2023,Los2024} and integrated quantum photonic architectures \cite{Rad2025,Larsen2025} are changing this paradigm.
In a recent experimental breakthrough \cite{Larsen2025}, generation of approximate single-mode Gottesman-Kitaev-Preskill (GKP) states \cite{Gottesman2001} by photon counting measurements on three modes of a four-mode Gaussian state was reported, with efficiencies of all three employed photon-number resolving detectors exceeding $96\%$, and reaching more than $99\%$ in the best case. With such technology advances, it is pertinent to focus on optimization of the state preparation schemes with respect to the generation probability.

\begin{figure}[b]
\centerline{\includegraphics[width=0.6\linewidth]{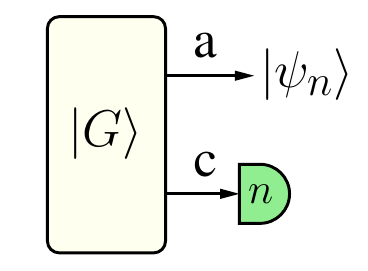}}
\caption{Conditional state preparation via Gaussian boson sampling \cite{Su2019,Hanamura2025}. Mode c of pure two-mode Gaussian state $|G\rangle$ is measured with photon number resolving detector. Detection of $n$  photons heralds preparation of state $|\psi_n\rangle$ in mode a.}
\label{figscheme}
\end{figure}

Very recently, single-mode state preparation via two-mode Gaussian boson sampling was investigated in detail in Ref. \cite{Hanamura2025}. As depicted in Fig.~1, a two-mode Gaussian state is generated and one mode is measured in Fock basis. A specific instance of this scheme is the generalized photon subtraction where the input two-mode Gaussian state is obtained by interference of two single-mode squeezed vacuum states at a beam splitter \cite{Takase2021,Tomoda2024,Takase2024,Takase2026}. As shown in Ref. \cite{Hanamura2025}, projection of mode c in Fig.~1 on Fock state $|n\rangle$ prepares the other mode a in a pure non-Gaussian state that can be expressed as
\begin{equation}
|\psi_n\rangle=\hat{U}_G(\hat{a}^\dagger+s_0\hat{a}+\delta_0)^n|0\rangle.
\label{psinintro}
\end{equation}
Here $\hat{U}_G$ denotes a fixed Gaussian unitary transformation that depends on the input state $|G\rangle$ but not on the measurement outcome $n$. The state $|\psi_n\rangle$ has stellar rank $n$  \cite{Chabaud2020,Chabaud2021,Lachman2019} and its non-Gaussian properties are fully specified by two so-called  control parameters $s_0$ and $\delta_0$ \cite{Hanamura2025}. The parameter $s_0$ can be considered real and non-negative while $\delta_0$ can be complex. Interestingly, the states (\ref{psinintro}) can very well  approximate important classes of states such as states with cubic nonlinear squeezing, or superpositions of coherent states \cite{Hanamura2025}. In Ref. \cite{Hanamura2025} optimization of the success probability of preparation of the state (\ref{psinintro}) by two-mode Gaussian boson sampling was discussed and numerical results were reported  for specific cases. Numerical optimization of the scheme in Fig.~1 for various specific target nonclassical states and  a limited amount of input squeezing was also reported in an earier work \cite{Mogyorosi2019}.

In this work, we further investigate the optimization of the success probability $P_n$ for generating the states (\ref{psinintro}) using the setup depicted in Fig.~1, and we show that this optimization can be carried out analytically. We derive analytical expressions for the maximum achievable success probability $P_n$ and study its asymptotic dependence on $n$. In particular, we show explicitly that when $s_0=0$ or $\delta_0=0$, the number of required state-generation trials, $P_n^{-1}$, scales only polynomially with $n$. Numerical calculations indicate that this favorable scaling also persists in the general case when both $\delta_0$ and $s_0$ are nonzero.

Our analysis is based on the Bargmann representation of quantum states of bosonic systems, which proves particularly well suited for studying the two-mode setup shown in Fig.~1. Since prior work \cite{Mogyorosi2019,Hanamura2025}  has firmly established that the class of states (\ref{psinintro}) encompasses a wide variety of important highly nonclassical states, we focus here specifically on maximizing the state-preparation probability. Accordingly, we treat the state parameters $s_0$ and $\delta_0$ as external inputs and analytically determine the maximum preparation probability of the state (\ref{psinintro}) for given values of $s_0$, $\delta_0$, and $n$.

The rest of the paper is organized as follows. In Sec~ II we introduce the formalism and review the Bargmann representation of pure quantum states. In Sec.~III we optimize the success probability of preparation $P_n$ of the state (\ref{psinintro}) and discuss the asymptotic dependence of $P_n$ on $n$. In Sec.~IV we consider the specific case of generation of $n$-photon added coherent states, which corresponds to the choice $s_0=0$. The quadrature squeezing required to achieve the maximum success probability is analyzed in Sec.~V. Effect of losses and noise on the considered conditional state preparation protocol is discussed in Sec~VI. In Sec. VII we  outline the possible extension of the optimization procedure to multimode schemes. Finally, Sec. VIII contains a brief summary and conclusions.

\section{Pure Gaussian states}
Any pure $M$-mode Gaussian state $|G\rangle$ can be written as
\begin{equation}
|G\rangle=Z \exp\left( \sum_{j,k=1}^M \hat{a}_{j}^\dagger A_{jk} \hat{a}_{k}^\dagger +\sum_{j=1}^M b_j \hat{a}_j^\dagger\right) |0\rangle.
\label{GMmode}
\end{equation}
Here $\hat{a}_j^\dagger$ denote creation operators, symmetric complex matrix $A$ determines the squeezing properties of the state, complex coefficients $b_j$ characterize the coherent displacement of the state, and $Z$ is a normalization constant.

The Gaussian function of creation operators that appears in Eq. (\ref{GMmode}) corresponds to the Bargmann representation of quantum states of bosonic systems \cite{Vourdas2006,Motamedi2025},
\begin{equation}
f(\bm{z})= e^{\frac{1}{2}|\bm{z}|^2 } \langle \bm{z}^\ast| G\rangle,
\end{equation}
where $|z\rangle=|z_1\rangle|z_2\rangle\cdots|z_M\rangle$ denotes $M$-mode coherent state with complex amplitudes $z_j$, and $\bm{z}=(z_1,z_2,\cdots,z_M)^T$ is a column vector. Hence 
\begin{equation}
 |\bm{z}|^2=\bm{z}^\dagger \bm{z}=\sum_{j=1}^M |z_j|^2.
 \end{equation}
 In this work we choose to work directly with functions of creation operators as in Eq. (\ref{GMmode}). 
 
The normalization factor $Z$ depends on both $A$ and $\bm{b}$ \cite{Motamedi2025}. The relation between the coherent displacements $\alpha_j$ of the state $|G\rangle$ and the parameters $b_j$ can be determined for instance by applying the inverse displacements $\hat{D}_j(-\alpha_j)=\hat{D}^\dagger_j(\alpha_j)$ to $|G\rangle$, requiring that the terms linear in $\hat{a}_j^\dagger$ disappear after such transformation,
\begin{equation}
\prod _{j=1}^M \hat{D}_j(-\alpha_j) |G\rangle = Z Z_D \exp\left( \sum_{j,k=1}^M \hat{a}_{j}^\dagger A_{jk} \hat{a}_{k}^\dagger \right) |0\rangle.
\label{Gdisplaced}
\end{equation}
After some algebra, one finds that
\begin{equation}
\bm{\alpha}=[I-4AA^\dagger]^{-1} (\bm{b}+2A \bm{b}^\ast)
\label{alpha}
\end{equation}
and
\begin{equation}
Z_D=\exp\left( \frac{1}{2}\bm{\alpha}^\dagger \bm{\alpha} -\bm{\alpha}^\dagger A\bm{\alpha}^{\ast}\right).
\end{equation}
Note that $A^\dagger=A^\ast$ because the matrix $A$ is symmetric.
The state in Eq. (\ref{Gdisplaced}) is an $M$-mode squeezed vacuum state. According to the Bloch-Messiah decomposition \cite{Braunstein2005}, it is possible to transform such state into a product  of $M$ single-mode squeezed vacuum states by a suitable $M$-mode passive linear Gaussian unitary transformation $\hat{U}_{\mathrm{IF}}$,
\begin{equation}
\hat{U}_{\mathrm{IF}} \prod _{j=1}^M \hat{D}_j(-\alpha_j) |G\rangle = Z Z_D \exp\left( \sum_{j}^M \tilde{A}_{jj} \hat{a}_{j}^{\dagger 2}  \right) |0\rangle.
\label{GIF}
\end{equation}
The linear interferometric coupling $\hat{U}_{\mathrm{IF}}$ induces  linear transformation of creation operators,
\begin{equation}
\hat{U}_{\mathrm{IF}}\hat{a}_j ^\dagger \hat{U}_{\mathrm{IF}}^\dagger= \sum_{k=1}^M V_{jk}\hat{a}_k^\dagger,
\label{alinear}
\end{equation}
where $V$ is an $M\times M$ unitary matrix. The linear transformation (\ref{alinear}) together with the vacuum stability condition $\hat{U}_{\mathrm{IF}} |0\rangle=|0\rangle$ implies Eq. (\ref{GIF}), where the transformed diagonal matrix $\tilde{A}$ reads
\begin{equation}
\tilde{A}=V^T A V.
\label{Adiagonalization}
\end{equation}
Any complex symmetric matrix $A$ can be diagonalized by the transformation (\ref{Adiagonalization}) and this is known as the Autonne–Takagi factorization \cite{Autonne1915,Takagi1925}. The diagonal elements $\tilde{A}_{jj}$ can be made real and nonnegative. It is now straightforward to connect Eq. (\ref{GIF}) with the product of $M$ single-mode squeezed vacuum states,
\begin{equation}
\prod _{j=1}^M (1-\mu_j^2)^{1/4} \exp\left( \frac{\mu_j}{2} \hat{a}_j^{\dagger 2}\right) |\bm{0}\rangle,
\end{equation}
where $\mu_j=\tanh r_j$ and $r_j$ is the squeezing constant of mode $j$. We can see that $\tilde{A}_{jj}=\mu_j/2$, i.e., the diagonalization (\ref{Adiagonalization})  reveals the single-mode squeezing constants. Observe that  $1-\mu_j^2$ are eigenvalues of matrix $ I-4 AA^\dagger$. Therefore, the following identity holds,
\begin{equation}
\prod _{j=1}^M (1-\mu_j^2)^{1/4}= [\det( I-4 AA^\dagger]^{1/4}.
\label{squeezingnormalization}
\end{equation}
With this expression at hand it is finally possible to specify the normalization factor $Z$ such that $ \langle G|G\rangle=1$ holds,
\begin{equation}
Z=\left[\det( I-4 AA^\dagger)\right]^{1/4} \exp\left(- \frac{1}{2}\bm{\alpha}^\dagger \bm{\alpha} +\bm{\alpha}^\dagger A\bm{\alpha}^{\ast}\right).
\end{equation}
Using Eq. (\ref{alpha}) it is possible to switch from the true displacements $\bm{\alpha}$ to parameters $\bm{b}$, which will be useful in what follows. The squeezing parameters $\mu_j$ must satisfy $|\mu_j|<1$. Consequently, the physicality condition can be formulated as a matrix inequality
\begin{equation}
I-4AA^\dagger >0,
\label{Aphysicality}
\end{equation}
which must be satisfied by any matrix $A$ that represents a physical Gaussian state $|G\rangle$.

\section{Conditional state preparation}

In this section we will  consider conditional generation of highly nonclassical single-mode states by photon counting measurements of one mode of pure two-mode Gaussian state, as depicted in Fig.~1.
A generic pure two-mode Gaussian state can be represented by Eq. (\ref{GMmode}) with $M=2$. As shown in Ref. \cite{Motamedi2025}, it is always possible to apply a suitable single-mode Gaussian unitary operation $\hat{U}_G^\dagger$ to the unmeasured mode which transforms the state (\ref{GMmode}) to the so-called core state. A key property of the core state is that projection of the measured mode onto Fock state $|n\rangle$ prepares the unmeasured mode in a finite superposition of Fock states  up to $|n\rangle$. The unitary $\hat{U}_G$ thus represents a Gaussian envelope that is independent of the measurement outcome $|n\rangle$ and can be removed to focus on the core non-Gaussian properties of the generated state. 

We shall call the unmeasured mode the signal mode and the measured mode the control mode, and we associate creation operators $\hat{a}^\dagger$ and $\hat{c}^\dagger$ with the signal and control modes, respectively. The core Gaussian state has the property that the terms in the exponent in Eq. (\ref{GMmode}) that depend only on the creation operator $\hat{a}^\dagger$ of the unmeasured mode vanish \cite{Motamedi2025}. A general pure two-mode core Gaussian state can thus be expressed as follows,
\begin{equation}
|G_\mathrm{core}\rangle= Z\exp \left( \frac{\mu}{2}\hat{c}^{\dagger 2} + \lambda \hat{a}^\dagger \hat{c}^\dagger+ \beta \hat{c}^\dagger\right) |0,0\rangle,
\label{Gcoretwomode}
\end{equation}
where $|0,0\rangle$ denotes the two-mode vacuum state and 
\begin{eqnarray}
|Z|^2&=&\sqrt{(1-\lambda^2)^2-\mu^2}  \nonumber \\
& & \times \exp\left[-\frac{(1-\lambda^2)|\beta|^2+\frac{1}{2}(\beta^2+\beta^{\ast^2})\mu }{(1-\lambda^2)^2-\mu^2}\right].
\end{eqnarray}
The parameters $\lambda$ and $\mu$ can be made real and nonnegative by suitable phase shifts applied to modes a and c, and we assume this in what follows.  On the other hand, $\beta$ can be complex. 

With the representation (\ref{Gcoretwomode}) it is straightforward to prove that projection of the control mode c onto Fock state $|n\rangle$ prepares the signal mode in state (\ref{psinintro}).
We make use of the identity 
\begin{equation}
e^{\kappa \hat{c}^\dagger \hat{a}} |0,0\rangle= |0,0\rangle
\end{equation}
to rewrite the state (\ref{Gcoretwomode}) equivalently as 
\begin{equation}
|G_\mathrm{core}\rangle= Z e^{\frac{\mu}{2}\hat{c}^{\dagger 2} +\beta \hat{c}^\dagger} e^{ \lambda \hat{a}^\dagger \hat{c}^\dagger} e^{\kappa \hat{c}^\dagger \hat{a}}  |0,0\rangle .
\end{equation}
Next we utilize the Baker-Campbell-Haussdorf identity 
\begin{equation}
e^{\hat{X}} e^{\hat{Y}}= e^{\hat{X}+\hat{Y}+\frac{1}{2}[\hat{X},\hat{Y}]}
\end{equation}
which holds when both $\hat{X}$ and $\hat{Y}$ commute with $[\hat{X},\hat{Y}]$. Specifically, we set $\hat{X}=\lambda \hat{a}^\dagger \hat{c}^\dagger$ 
and $\hat{Y}=\kappa \hat{c}^\dagger \hat{a}$ to obtain
\begin{equation}
|G_\mathrm{core}\rangle=Z e^{\frac{\mu}{2}\hat{c}^{\dagger 2} +\beta \hat{c}^\dagger} e^{ \lambda \hat{a}^\dagger \hat{c}^\dagger+\kappa \hat{c}^\dagger \hat{a}-\frac{\kappa \lambda}{2}\hat{c}^{\dagger 2}}  |0,0\rangle.
\end{equation}
This expression simplifies when we set $\kappa=\mu/\lambda$,
\begin{equation}
|G_\mathrm{core}\rangle= Z\exp\left[ \lambda \hat{a}^\dagger \hat{c}^\dagger+\frac{\mu}{\lambda} \hat{c}^\dagger \hat{a} +\beta \hat{c}^\dagger \right]  |0,0\rangle.
\end{equation}
Finally, we introduce the real parameter $s_0$ and complex parameter $\delta_0$,
\begin{equation}
\mu= \lambda^2 s_0, \qquad \beta =\delta_0 \lambda,
\end{equation}
which results in
\begin{equation}
|G_\mathrm{core}\rangle= Z\exp\left[ \lambda \hat{c}^\dagger \left( \hat{a}^\dagger + s_0 \hat{a} +\delta_0\right) \right]  |0,0\rangle.
\label{Gcorefinal}
\end{equation}
When we expand the exponential operator in Taylor series, we immediately find that the conditionally generated state in mode a when mode c is projected into Fock state $|n\rangle$  reads
\begin{equation}
|\psi_n\rangle_a =  \left( \hat{a}^\dagger + s_0 \hat{a} +\delta_0\right)^n |0\rangle.
\label{psin}
\end{equation}
Moreover, we can directly write down a formula for the success probability of preparation of this state,
\begin{equation}
P_n= \frac{\lambda^{2n}}{n!} |Z|^2 \langle \psi_n|\psi_n\rangle ,
\label{Pn}
\end{equation}
where $|Z|^2$ depends on $s_0$, $\delta_0$ and $\lambda$. Note that $\lambda$ is a free parameter that can be optimized to maximize the success probability $P_n$ \cite{Hanamura2025}. The optimal value of $\lambda$ can be found from the extremality condition
\begin{equation}
\frac{d P_n}{\lambda}=0.
\label{lambdaextremal}
\end{equation}
As we now show, this leads to polynomial equation for $\lambda^2$. 

\begin{figure}[t]
\centerline{\includegraphics[width=0.8\linewidth]{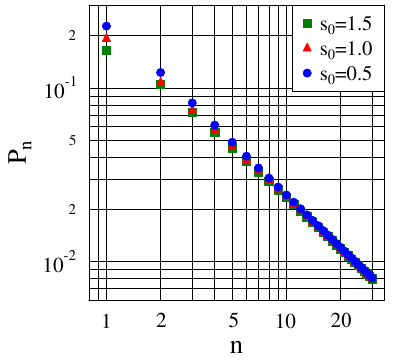}}
\caption{Dependence of the  heralding probability $P_n$ on detected number of photons $n$ is ploted for three different values of control parameter $s_0$, and $\delta_0=0$.} 
\label{figPnparity}
\end{figure}

Let us first consider the case $\delta_0=0$. In such case the generated state (\ref{psin}) has a well defined parity in Fock space, which is given by the parity of $n$ \cite{Korolev2024,Korolev2025},
\begin{equation}
|\psi_n\rangle_A=\sum_{m=0}^{ \left\lfloor\frac{n}{2}  \right\rfloor} \frac{s_0^m n!}{2^m m!\sqrt{(n-2m)!}} |n-2m\rangle.
\label{psinparity}
\end{equation}
\tcr{The states with well defined photon number parity which can be generated with the analyzed scheme include high-quality odd and even cat-like states \cite{Hanamura2025}, i.e.,  high-fidelity approximations of superpositions of coherent states }
\begin{equation}
\tcr{|\alpha_{\pm}\rangle=\frac{1}{\sqrt{2(1\pm e^{-2|\alpha|^2})}}(|\alpha\rangle\pm|-\alpha\rangle),}
\end{equation}
\tcr{ Such states  represent essential  resource for 
quantum computing with qubits encoded into  superpositions of coherent states \cite{Ralph2003}. }

\tcr{For target states with well defined photon number parity (\ref{psinparity})  we have}  $\beta=0$ and the expression for $P_n$ simplifies,
\begin{equation}
P_n=\langle \psi_n|\psi_n\rangle \frac{\lambda^{2n}}{n!}\sqrt{(1-\lambda^2)^2-s_0^2\lambda^4}.
\label{Pnparitysimplified}
\end{equation}
The extremal equation (\ref{lambdaextremal}) yields quadratic equation for $\lambda^2$, whose two roots read
\begin{equation}
\lambda^2_{1,2}=\frac{2n+1\pm\sqrt{1+4n(n+1)s_0^2}}{2(1-s_0^2)(n+1)} .
\end{equation}
It turns out that the root with the minus sign corresponds to the optimal value of $\lambda^2$ that maximizes $P_n$. 

\begin{figure}[t]
\centerline{\includegraphics[width=0.8\linewidth]{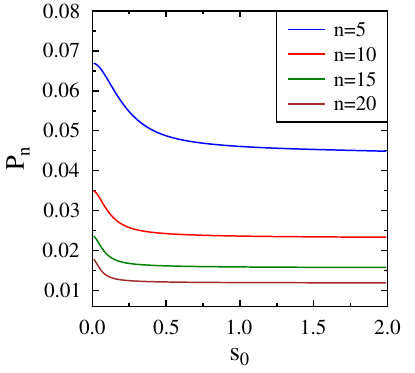}}
\caption{Dependence of the  heralding probability $P_n$ on the control parameter $s_0$  is plotted for four different values of $n$, and $\delta_0=0$.}
\label{figPnparitysdependence}
\end{figure}

The dependence of $P_n$ on $n$ and $s_0$ is illustrated in Fig.~\ref{figPnparity} and Fig.~\ref{figPnparitysdependence}, respectively. We can observe that $P_n$ decreases only polynomially with increasing $n$, and the log-log plot in Fig.~\ref{figPnparity}  suggests scaling $P_n \propto n^{-1}$. Furthermore, Fig.~\ref{figPnparitysdependence} shows that the maximum achievable $P_n$ depends only weakly on $s_0$ for the range of parameters considered.

\begin{figure*}
\centerline{\includegraphics[width=0.98\linewidth]{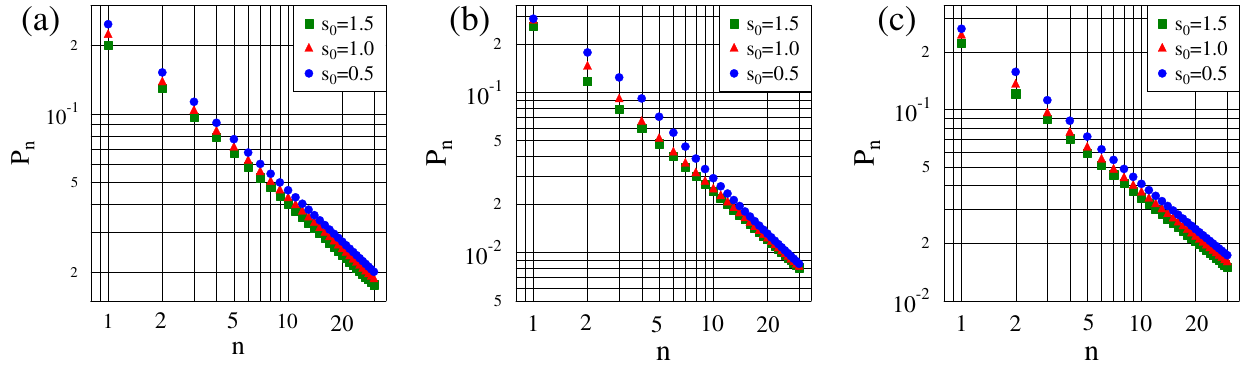}}
\caption{The maximum heralding probability $P_n$ is plotted as a function of $n$ for  $\delta_0=1$ (a), $\delta_0=i$ (b), and $\delta_0=e^{i\pi/4}$ (c). In each case, results for three different values of the oher control parameter $s_0$ are plotted, $s_0=0.5$ (blue circles), $s_0=1$ (red triangles), and $s_0=1.5$ (green squares).}
\end{figure*}

Let us investigate asymptotic behavior of $P_n$ in more detail. We shall assume that $s_0$ is positive. In the large $n$ limit we have
\begin{equation}
\lambda^2 \approx \frac{1}{1+s_0}\left( 1 -\frac{1}{2n}\right),
\end{equation}
hence
\begin{equation}
\lambda^{2n} \sqrt{(1-\lambda^2)^2-s_0^2\lambda^4} \approx  \frac{1}{\sqrt{n}}\frac{e^{-1/2}}{(1+s_0)^n} \sqrt{\frac{s_0}{1+s_0}}.
\label{lambdaasymptotics}
\end{equation}
Assuming even $n$, the norm of the state (\ref{psinparity}) can be lower bounded as follows
\begin{eqnarray}
\langle \psi_n|\psi_n\rangle & =& n! \sum_{m=0}^{\frac{n}{2}} \frac{s_0^{2m} n!}{2^{2m}(m!)^2 (n-2m)!} \nonumber  \\
&\geq &  n! \sum_{m=0}^{\frac{n}{2}} \frac{s_0^{2m} n!}{2\sqrt{ m+1}(2m)! (n-2m)!} \nonumber  \\
&\geq &   \frac{n!}{2\sqrt{(n/2+1)}}\sum_{m=0}^{\frac{n}{2}}  {n \choose 2m}s_0^{2m} \nonumber  \\
&= &  \frac{n!}{2\sqrt{(n/2+1)}} \sum_{m=0}^{n}  {n \choose m}\frac{1}{2}\left[s_0^{m} + (-s_0)^m\right] \nonumber  \\
& =& \frac{n! }{2\sqrt{(2n+4)}} \left[(1+s_0)^n+(1-s_0)^n\right].
\label{psinorminequality}
\end{eqnarray}
The first inequality in Eq. (\ref{psinorminequality}) follows from the inequality
\begin{equation}
2^{2m} m! \, m! \leq 2\sqrt{ (m+1)} (2m)!.
\end{equation}
The second inequality is obtained by replacing $m+1$ with $n/2+1$ in the denominator. 
If we combine together Eqs. (\ref{Pnparitysimplified}), (\ref{lambdaasymptotics}) and (\ref{psinorminequality}) we find out that the factorial $n!$ and the exponential terms $(1+s_0)^n$ cancel out and $P_n^{-1}$ asymptotically scales polynomially with $n$,  $P_n \propto n^{-1}$. This scaling is fully consistent with the exact resuts plotted in  Fig.~\ref{figPnparity}.

To obtain additional insight, we consider the  point $s_0=1$, where the norm of $|\psi_n\rangle$ can easily be evaluated analytically. Specifically, the optimal parameter  $\lambda^2$ reads $\lambda^2=n/(2n+1)$, and 
\begin{equation}
\langle \psi_n|\psi_n\rangle=2^{n} \langle 0 |\hat{x}^{2n}|0\rangle=\frac{2^n}{\sqrt{\pi}} \Gamma \left(n+\frac{1}{2}\right).
\end{equation}
Here $\hat{x}=(\hat{a}+\hat{a}^\dagger)/\sqrt{2}$ is the quadrature operator, and $\Gamma(x)$ denotes the Euler Gamma function. For large $n$, we can approximate the Gamma function using Stirling's formula which yields
\begin{equation}
P_n \approx \frac{e^{-1/2}}{\sqrt{2\pi} n},
\end{equation}
valid at $s_0=1$. This approximate formula is in excellent agreement with the exact results plotted in Fig.~\ref{figPnparity}. Another case that allows exact treatment is the generation of (squeezed) Fock states, $s_0=0$ \cite{Korolev2024b}. We get $\lambda^2=n/(n+1)$ and  
\begin{equation}
P_n=\frac{n^n}{(n+1)^{n+1}} \approx \frac{1}{e n},
\end{equation}
where the  approximation holds in the asymptotic large $n$ limit.

Let us now consider the general situation when both control parameters $s_0$ and $\delta_0$ are nonzero. The extremal equation (\ref{lambdaextremal}) becomes a fourth-order polynomial equation for $\lambda^2$,
\begin{eqnarray}
& & n-(1+|\delta_0|^2+4n)\lambda^2 +(1-s_0^2)^2(1+n)\lambda^8 \nonumber \\
& &+\left[2|\delta_0|^2+4n+3-s_0^2+2n(1-s_0^2)-(\delta_0^2+\delta_0^{\ast 2})s_0\right]\lambda^4  \nonumber \\
& &-\left[(1-s_0^2)(3+4n)+|\delta_0|^2(1+s_0^2)-(\delta_0^2+\delta_0^{\ast 2})s_0\right]\lambda^6=0.  \nonumber \\
\label{lambdafour}
\end{eqnarray}
Note that  we seek a positive root $\lambda^2$ that satisfies the physicality condition $\lambda^2<1/(1+s_0)$. The roots of the equation  (\ref{lambdafour}) can be expressed analytically, but the resulting formulas are very lengthy and we do not reproduce them here. Instead, in Fig.~4 we plot the resulting dependence of the maximum achievable  $P_n$ on $n$ for several diferent combinations of $s_0$ and $\delta_0$. We can see that the scaling of $P_n^{-1}$ with $n$ is again polynomial, of the form $P_n \propto n^{-\gamma}$. The value of $\gamma$ generally depends on $s_0$ and $\delta_0$. To illustrate this, we in the next section investigate in more detail the case $s_0=0$ and $\delta_0 \neq 0$.

\section{Photon-added coherent states}

In this section we shall investigate the probability of conditional generation of $n$-photon-added coherent states \cite{Agarwal1991,Zavatta2004,Fadrny2024},
\begin{equation}
|\phi_n\rangle= \hat{a}^{\dagger n}|\alpha\rangle.
\end{equation}
\tcr{The photon added coherent states are an example of states that exhibit cubic nonlinear squeezing \cite{Kala2025}. The cubic squeezing occurs when the variance of the operator  
$\hat{p}+\hat{x}^2$ is reduced below the minimum achievable with Gaussian states \cite{Kala2022}. States exhibiting strong cubic squeezing can be generated with the considered stetup in Fig.~1  \cite{Hanamura2025}. The nonlinearly squeezed states can serve as a valuable resource for implementation of qubic  phase gates in optical quantum computing \cite{Marek2011,Miyata2016,Sakaguchi2023}.}

The photon-added coherent states can be equivalently expressed as \cite{Agarwal1991}
\begin{equation}
|\phi_n\rangle= \hat{D}(\alpha)(\hat{a}^{\dagger} +\alpha^\ast)^n |0\rangle
\label{phinadded}
\end{equation}
which exactly agrees with Eqs. (\ref{psinintro}) and (\ref{psin}) with $s_0=0$ and  $\delta_0=\alpha^\ast$. The norm of the state (\ref{phinadded}) can be expressed in terms of Laguerre polynomials $L_n(x)$,
\begin{equation}
\langle \phi_n|\phi_n\rangle=n! L_n(-|\alpha|^2).
\end{equation}
Since $s_0=0$, the formula for $P_n$ simplifies considerably,
\begin{equation}
P_n=(1-\lambda^2)\lambda^{2n}  L_n(-|\alpha|^2) \exp\left(-\frac{\lambda^2 }{1-\lambda^2}|\alpha|^2\right).
\label{Pnadded}
\end{equation}
Consequently, the extremal equation (\ref{lambdaextremal}) reduces again to a quadratic equation. Its root which corresponds to the optimal squeezing $\lambda^2$ reads
\begin{equation}
\lambda^2=\frac{1}{2(n+1)}\left[2n+1+|\alpha|^2 -\sqrt{(1+|\alpha|^2)^2+4n|\alpha|^2}\right].
\end{equation}
In the large $n$ limit we obtain 
\begin{equation}
\lambda^{2n} \approx e^{-\sqrt{n}|\alpha|-1/2}, \qquad e^{-\frac{\lambda^2 }{1-\lambda^2}|\alpha|^2} \approx e^{-\sqrt{n}|\alpha|} e^{(1+|\alpha|^2)/2},
\label{addedasymptotics}
\end{equation}
and 
\begin{equation}
\lambda^2 \approx 1-\frac{|\alpha|}{\sqrt{n}}.
\end{equation}
Asymptotic  behavior of Laguerre polynomials for large $n$  and negative arguments is described by Perron's formula  \cite{Szego1979,Alvarez2004},
\begin{equation}
L_n(-x)=\frac{e^{-x/2} e^{2\sqrt{nx}}}{2\sqrt{\pi}(nx)^{1/4}} \left[1+O\left(n^{-1/2}\right)\right].
\label{LnPerron}
\end{equation}
If we insert the asymptotic expressions (\ref{addedasymptotics}) and (\ref{LnPerron}) into the formula (\ref{Pnadded}) for $P_n$, we get
\begin{equation}
P_n \approx \frac{\sqrt{|\alpha|}  }{2\sqrt{\pi} n^{3/4}} .
\label{Pnaddedasymptotic}
\end{equation}
Interestingly, for the class of states with $s_0=0$ we get slightly different scaling of $P_n$ with $n$ than for the class $\delta_0=0$, namely $P_n\propto n^{-3/4}$.
Explicit calculations based on the exact formula (\ref{Pnadded}) confirm the validity of the aymptotic formula (\ref{Pnaddedasymptotic}), although for small $|\alpha|$  the  asymptotic values are approached only for extremely large $n$.

\begin{figure}[b]
\centerline{\includegraphics[width=0.9\linewidth]{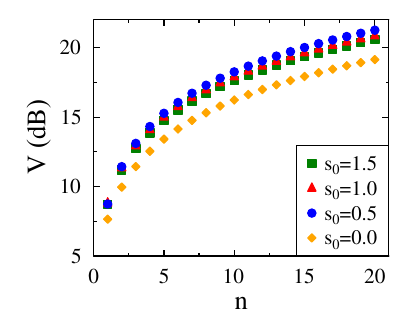}}
\caption{Single-mode squeeezing $V$ that is required to achieve the maximum preparation probability $P_n$ is plotted in dependence on $n$ for $\delta_0=0$  and four different values of  control parameter $s_0$.} 
\label{figVs0}
\end{figure}

\section{Required quadrature squeezing}

The quadrature squeezing required for achieving the maximum generation probability is a critical resource. The optimal core Gaussian  state $|G_{\mathrm{core}}\rangle$ can be prepared by interference of two single-mode squeezed coherent states at a beam splitter \cite{Braunstein2005,Mogyorosi2019}. The squeezing constants $r_{\pm}$ of these constituent single-mode states can be  determined from the eigenvalues of the matrix $AA^\dagger$, cf. Eq. (\ref{squeezingnormalization}) in Sec.~II and the related discussion. For the two-mode core Gaussian state (\ref{Gcorefinal}) we obtain
\begin{equation}
\tanh^2 r_{\pm}= \frac{1}{2}\lambda^2\left(2+s_0^2\lambda^2\pm s_0\lambda \sqrt{4+s_0^2\lambda^2}\right).
\end{equation}
Note that the squeezing does not depend on $\delta_0$ since this latter parameter is fully controlled by coherent displacements that do not modify the squeezing. For $s_0=0$ we get  $\tanh^2 r_{\pm}=\lambda^2$ and the core Gaussian state becomes a (possibly coherently displaced) two-mode squeezed vacuum state. 

\begin{figure}[t]
\centerline{\includegraphics[width=0.99\linewidth]{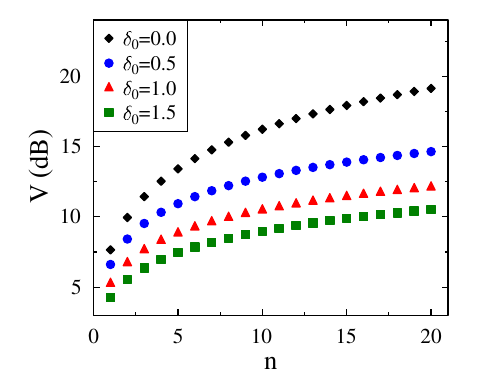}}
\caption{Single-mode squeezing $V$ that is required to achieve the maximum preparation probability $P_n$ is plotted in dependence on $n$ for $s_0=0$  and four different values of  control parameter $\delta_0$.} 
\label{figVdelta0}
\end{figure}

For fixed $s_0$ and $\delta_0$, the squeezing required to achieve the maximum preparation probability increases with the number of detected heralding photons $n$. To assess the typical required level of squeezing, we plot the required squeezing variance $V$ in dB as a function of $n$ for two illustrative cases in Figs.~\ref{figVs0} and \ref{figVdelta0}. The squeezing variance in dB is defined as 
\begin{equation}
V=10 \log_{10}e^{2r_{+}} \approx 8.686 r_{+}.
\end{equation}

Figure~\ref{figVs0} shows the dependence of $V$ on $n$ for several different values of $s_0$ and $\delta=0$, which is the same choice of parameters as in Fig.~2. We can observe that the optimal squeezing increases with $n$ as expected and it only weakly depends on $s_0$. For $n=20$ the optimal squeezing is around $21$~dB.  In Fig.~\ref{figVdelta0} we plot the dependence of $V$ on $n$ for $s_0=0$ and several values of $\delta_0$, which corresponds to optimal preparation of $n$-photon added coherent states (\ref{phinadded}) with $|\alpha|=\delta_0$. 
In Fig.~\ref{figVdelta0} we can observe that the squeezing required for optimal performance decreases with increasing $\delta_0$. For $n=20$, optimal $V$ is less than $15$~dB for $\delta_0\gtrsim 0.5$. To obtain more insight into the typical scaling of $V$ with $n$, it is instructive to consider conditional preparation of Fock states. We have $s_0=0$,  the optimal $\lambda$ is given by $\lambda^2=n/(n+1)$, and
\begin{equation}
e^{2 r_{+}}=(n+1)\left[1+\sqrt{\frac{n}{n+1}}\right]^2.
\end{equation}
The squeezing variance $e^{2r_+}$ grows linearly with $n$, hence $V$ grows logarithmically with $n$ in dB units.

Currently, the maximum directly observed quadrature squeezing is $15$ dB \cite{Vahlbruch2016}, while in the recent Xanadu experiment on preparation of GKP states \cite{Larsen2025} squeezed states with 10 dB and 8 dB of squeezing (before loss), were generated and utilized. In experiments, the observable squeezing is mainly reduced by losses and phase noise \cite{Ha2026}. Losses resulting in effective total intensity transmittance $\eta$  bound any observable quadrature squeezing at $ -10\log_{10} \eta$. Phase noise mixes together squeezed and anti-squeezed quadratures and its detrimental effect is therefore stronger for higher initial squeezing.

\section{Losses and noise}

\begin{figure}[t]
\centerline{\includegraphics[width=\linewidth]{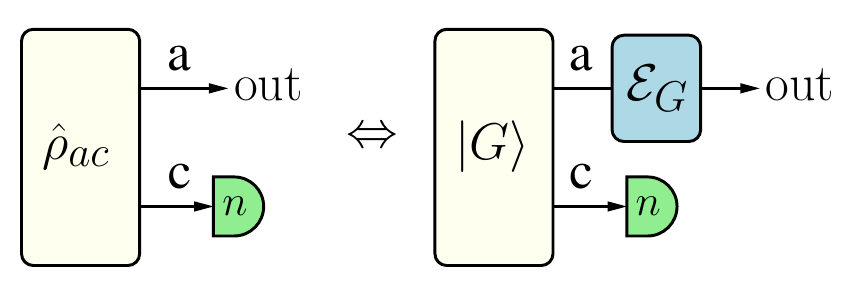}}
\caption{Equivalence of two-mode Gaussian quantum states. Any mixed two-mode Gaussian state $\hat{\rho}_{\mathrm{ac}}$ can be represented as a pure two-mode (core) Gaussian state $|G\rangle_{ac}$ followed by a single-mode Gaussian quantum channel $\mathcal{E}_G$ that acts on  mode a only.} 
\label{figschemeEG}
\end{figure}

In experimental practice, losses and noise will almost inevitably affect the quantum state generation process. Interestingly, for the two-mode scheme considered in the present paper, all noise and losses can be represented by a single-mode Gaussian quantum channel $\mathcal{E}_G$ that acts on the signal mode a \cite{Motamedi2025}. More specifically,  as illustrated in Fig.~\ref{figschemeEG}, any two-mode mixed Gaussian quantum state $\hat{\rho}_{\mathrm{ac}}$ can be obtained from certain pure two-mode (core) Gaussian  state $|G\rangle$ by sending the mode a through some single-mode noisy Gaussian quantum channel $\mathcal{E}_G$ \cite{Motamedi2025}. If the noise and losses present in the photon counting measurement on mode c can be incorporated into the Gaussian state $\rho_{ac}$, then  the equivalence established in Fig~\ref{figschemeEG} implies that the conditionally generated mixed state in mode a is obtained by sending a conditionally generated pure (core) non-Gaussian state (\ref{psin}) through the Gaussian noisy channel $\mathcal{E}_G$.

The Gaussian quantum channel $\mathcal{E}_G$ is fully described by two matrices $S_a$ and $M_a$ which govern the transformation of the covariance matrix $\gamma_a$ of mode a,
\begin{equation}
\gamma_a\rightarrow S_a \gamma_a S_a^T+M_a,
\label{Gchannel}
\end{equation}
 and by coherent displacement which is added in the channel. Since coherent displacements are local operations that do not affect the noise of the state, we focus here on the transformation of the covariance matrix.
The matrices $S_a$ and $M_a$ must satisfy the condition 
\begin{equation}
 M_a+ i\Omega-S_a i\Omega S_a^T \geq 0,
\label{cpcondition}
\end{equation}
where 
\begin{equation}
\Omega= \left(
\begin{array}{cc}
 0 & 1 \\
 -1 & 0 
\end{array}
\right)
\end{equation}
is the single-mode symplectic form. Note that the covariance matrices are normalized such that the covariance matrix of vacuum is equal to the identity matrix.  The inequality (\ref{cpcondition}) ensures that the quantum channel (\ref{Gchannel}) is a Gaussian completely positive map.

 Explicit  procedure for decomposition of a mixed Gaussian state $\hat{\rho}$ into the pure core Gaussian state and a single-mode Gaussian quantum channel is provided in Ref. \cite{Motamedi2025}, based on Bargmann parametrization.  For completeness,  we outline here the main steps when the calculations are performed directly with the covariance matrix of the state and we omit the coherent displacements, since they do not affect the noise properties of the state. Consider the mode-wise decomposition of the covariance matrix of $\hat{\rho}_{ac}$,
\begin{equation}
\gamma_{\mathrm{ac}}=\left(
\begin{array}{cc}
\gamma_\mathrm{a} & \sigma \\
\sigma^T & \gamma_\mathrm{c}
\end{array}
\right).
\label{gammaac}
\end{equation}
In what follows we assume that the two-mode Gausisan state $\hat{\rho}_{\mathrm{ac}}$ with covariance matrix (\ref{gammaac}) is entangled, which implies that $\det \sigma <0$ \cite{Simon2000}.
It is convenient to introduce a notation for diagonal symplectic matrix that describes the single-mode squeezing operation,
\begin{equation}
S_{\mathrm{sq}}(r)=\left(
\begin{array}{cc}
e^{r} & 0 \\
0 & e^{-r}
\end{array}
\right).
\end{equation}
In order to perform the decomposition depicted in Fig.~\ref{figschemeEG}, we first find a single-mode symplectic matrix $S_c$ that generates $\gamma_c$ from thermal state,  $\gamma_c=\sqrt{\det \gamma_c} S_c S_c^T$. This can be easily accomplished by introducing the decomposition $S_c= O_cS_{\mathrm{sq}}(r_c)$, where $O_c$ is an orthogonal matrix with $\det O_c=1$, and finding diagonalization of $\gamma_c$.
The matrix $S_a$ can then be constructed as
\begin{equation}
S_a= \frac{1}{\sinh(2r)}\sigma (S_c^T)^{-1} \Sigma S_{\mathrm{sq}}^{-1}(r_a)
\end{equation}
 where $\cosh(2r)=\sqrt{\det\gamma_c}$,  $\tanh r_a= (\tanh r)^2 \tanh r_c$, $r>0$, and $\Sigma=\mathrm{diag}(1,-1)$  Finally, 
 \begin{equation}
 M=\gamma_a-\sqrt{\det \gamma_c}S_a S_{\mathrm{sq}}^2(r_a)S_a^T.
 \end{equation}
 
 The corresponding pure two-mode core Gaussian state $|G\rangle$ is obtained from two-mode squeezed vacuum state with squeezing constant $r$ by applying the single-mode squeezing operation with squeezing constant $r_a$  to mode a, and a single-mode unitary Gaussian operation described by symplectic matrix $S_c$ to mode c, respectively. Covariance matrix of this pure core Gaussian state is given by
 \begin{equation}
\gamma_{\mathrm{core}}=\left(
\begin{array}{cc}
\cosh(2r) S_{\mathrm{sq}}^2(r_a) &  \sinh(2r) S_{\mathrm{sq}}(r_a)\Sigma S_c^T \\[1mm]
\sinh(2r)S_c \Sigma S_{\mathrm{sq}}(r_a)&  \cosh(2r)S_c S_c^T
\end{array}
\right).
 \end{equation}
Note that the local squeezing $S_{\mathrm{sq}}(r_a)$ of mode a is tuned such that if  mode c is projected on vacuum state then mode a is prepared in vacuum state, as expected for core Gaussian state.

\begin{figure}[t]
\centerline{\includegraphics[width=\linewidth]{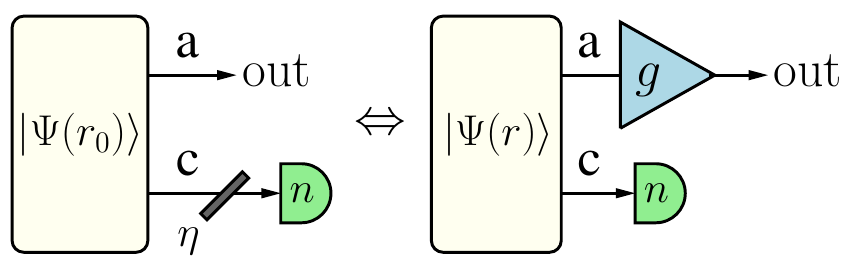}}
\caption{Illustration of the effect of losses in the heralding mode c. A two-mode squeezed vacuum state $|\Psi(r_0)\rangle$ followed by losses in mode c is equivalent to a two-mode squeezed vacuum state with modified squeezing constant $r$ followed by phase insensitive amplifier with gain $g$ acting on mode a. Analytical expressions for $r$ and $g$ are provided in the main text.} 
\label{figschemeloss}
\end{figure}

As a simple yet illustrative example let us consider the effect of losses in the heralding mode c on conditional preparation of Fock states or, more generally, $n$-photon added coherent states, see Fig.~\ref{figschemeloss}.
In this case, the optimal Gaussian core state is the two-mode squeezed vacuum state, possibly coherently displaced. We therefore assume that $\hat{\rho}_{ac}$ is obtained from (possibly coherently displaced) two-mode squeezed vacuum with squeezing constant $r_0$ by sending mode c through a lossy channel with transmittance $\eta$. This yields
\begin{equation}
\gamma_{ac}=
\left(
\begin{array}{cc}
\cosh(2r_0) I & \sqrt{\eta} \sinh(2r_0) \Sigma \\[1mm]
\sqrt{\eta} \sinh(2r_0) \Sigma & [\eta \cosh(2r_0)+1-\eta]I
\end{array}
\right).
\label{cmlosses}
\end{equation}
The losses in mode c can account for losses in state preparation, transmission of mode c from the source to the detector, and non-unit detection efficiency of the photon counting detector.

Following the above outlined formalism we find that a Gaussian state with covariance matrix  (\ref{cmlosses}) can be equivalently obtained from a two-mode squeezed vacuum state with modified squeezing constant 
$r$ given by
\begin{equation}
\sinh^2(r)=\eta\sinh^2(r_0),
\end{equation}
whose mode a is sent through a phase insensitive linear amplifier with amplification gain 
\begin{equation}
g=\frac{\cosh r_0}{\sqrt{1+\eta \sinh^2 r_0}}.
\end{equation}
Recall that the phase-insensitive amplification channel is specified by matrices $S_a=gI$ and $M_a=(g^2-1)I$, $g \geq 1$. The equivalence between losses in mode c and phase insensitive noisy amplification of mode a is illustrated in Fig.~\ref{figschemeloss}.
The added thermal noise increases with increasing amplification gain $g$, i.e. with increasing two-mode squeezing constant $r_0$ and decreasing transmittance $\eta$. For large $r_0$ the gain saturates at $g\approx 1/\sqrt{\eta}$. Since the required optimal squeezing increases with $n$, the sensitivity of the optimal protocol to losses will increase with $n$.

\tcr{Besides  losses, the photon counting can be affected by noise such as dark counts. The relative number of incorrect heralding events due to dark counts can be estimated as $R_D\tau/ P_n$, where $\tau$ is the width of the detection window, $R_D$ is the dark count rate, and $P_n$ denotes the heralding probability. The width of the detection window $\tau$ can typically range from a few nanoseconds to a few tens of nanoseconds.  Current state of the art experiments \cite{Eaton2023,Larsen2025} employ transition-edge sensors (TES)  \cite{Lita2008,Fukuda2011} that exhibit negligible dark counts, very high detection efficiency,  and true photon number resolving capability. 
Alternatively, multiplex of superconducting nanowire single-photon detectors (SNSPDs) can be utilized to achieve approximate photon number resolution \cite{Cheng2023}. The dark count rates of SNSPDs are usually  below $100$ cps.}

\tcr{ In addition to  dark counts, the cross talks between the detection channels or the counts caused by stray light can degrade the precise photon number resolving capability of the detector. In the recent experiment by Xanadu on the generation of GKP states \cite{Larsen2025}, this noise was suppressed by discarding the 20\% of noisiest outcomes of each of the photon number resolving TES detectors. When arrays of detectors that only distinguish the presence and absence of photons are employed for approximate photon counting, the input  Fock state  $|n\rangle$ can yield less than $n$ clicks, which can result in inaccurate detection. For example, the probabilities that the Fock states $|n\rangle$ and $|n+1\rangle$ will generate $n$ clicks of a multiplexed detector consisting of $M$ binary click detectors read
\begin{equation}
C_{n,n}=\frac{M!}{(M-n)!M^n}, \qquad C_{n,n+1}= C_{n,n} \frac{n(n+1)}{2M}.
\end{equation}
In experiments with strongly squeezed light the probabilities that the state contains $n$ or $n+1$ photons can be comparable. To suppress the detrimental effect of cross talks, $C_{n,n+1}\ll C_{n,n}$ must hold, which implies the condition $M \gg n^2$, hence a very large number of  detection channels $M$ in the multiplex must be utilized to faithfully detect higher photon numbers $n$.}

\section{Extension to multimode schemes}

In our work we have utilized the concept of core Gaussian states and the Bargmann representation which naturally lead  to an efficient and simple parametrization of the studied states. In particular, the parameter $\lambda$ straightforwardly emerged as a free parameter that can be optimized. As pointed out in Ref. \cite{Hanamura2025}, the existence of such free parameter is a consequence of the independence of the conditionally generated state in mode a on Gaussian transformations of mode c which commute with the photon number operator in that mode, $\hat{n}_c=\hat{c}^\dagger \hat{c}$ . This includes unitary phase shifts $e^{i\phi \hat{n}_c}$ but also non-unitary operations corresponding to imaginary phase shift. The resulting operation $\lambda^{\hat{n}_{c}}$ can be either noiseless attenuation \cite{Micuda2012,Nunn2021} or noiseless amplification \cite{Ralph2009}, depending on the value of $\lambda$. If we represent the state as a function of creation operators acting onto vacuum, the transformation of $|G_{\mathrm{core}}\rangle$ by  $\lambda^{\hat{n}_{c}}$ results in a simple rescaling of the creation operator, $\hat{c}^\dagger \rightarrow \lambda\hat{c}^\dagger$, see the exponent in Eq. (\ref{Gcorefinal}).

This can be straightforwardly generalized to multimode scenario. Let us assume that the $M$ modes are split to $N=M-K$ unmeasured modes and $K$ modes measured each in Fock basis, and conditioning on a particular sequence of numbers of detected photons $(n_1,n_2,\cdots,n_K)$ is performed. In such case we can consider linear scaling of creation operator of each  measured mode \cite{Hanamura2025}. We  collect the scaling factors into a diagonal matrix $\Lambda=\mathrm{diag}(1,1,\cdots,1,\lambda_1,\lambda_2,\cdots,\lambda_{K})$. The corresponding transformation of the $M$-mode Gaussian state $|G\rangle$  in Eq. (\ref{GMmode}), which does not change the conditionally generated state in the first $M-K$  modes, can be succinctly expressed as transformation of matrix $A$ and vector $\bm{b}$,
\begin{equation}
A\rightarrow \Lambda A \Lambda,\qquad \bm{b}\rightarrow \Lambda \bm{b},
\label{Htransform}
\end{equation}
together with the corresponding change of the normalization factor, to keep the state properly normalized. The coefficients $\lambda_j$ are limited by the physicality condition (\ref{Aphysicality}). The simplicity of Eq. (\ref{Htransform}) suggests that the formalism employed in this work can be useful and efficient also for study of more complex multimode conditional state preparation schemes. 
The extremal equations for optimal $\lambda_j$ which maximize the preparation probability $P_n$ will have the form of a system of polynomial equations and only numerical optimization will be possible in general. Note also, that  the total number of parameters that characterize the core Gaussian state scales quadratically with the number of modes and in the multimode case the set of free parameters that can be potentially optimized can  thus include also other parameters than $\lambda_j$.

\section{Summary and Conclusions}

In summary, we have investigated the probability of generation of nonclassical single-mode states of light by photon counting measurement on one mode of a two-mode entangled pure Gaussian state. We have shown that the maximum heralding probability can be calculated analytically and simple formulas were obtained for the special cases when one of the control parameters $s_0$ or $\delta_0$ is equal to zero. We have investigated asymptotic scaling of the heralding probability and we have observed that $P_n^{-1}$ scales polynomially with $n$. Even for $n$ as large as $20$ the achievable  heralding probabilities are of the order of $10^{-2}$, which suggests that the states can be experimentally generated with sufficiently high repetition rate. For given fixed control parameters $s_0$ and $\delta_0$ the squeezing required to maximize the state generation probability $P_n$ increases with $n$ and the available squeezing may in practice limit the maximally achievable $P_n$.  Nevertheless, our work provides important analytical benchmarks and shows what is the maximum potentially achievable $P_n$. 
We have also briefly discussed the effect of noise and losses in the considered state generation protocol. Following the results reported in Ref. \cite{Motamedi2025}, we have pointed out that a scheme with noisy two-mode Gaussian state is equivalent to a scheme with certain pure two-mode Gaussian state, followed by propagation of the conditionally generated state through some noisy Gaussian channel. 

In the present work we take the control parameters $s_0$ and $\delta_0$ as given inputs and we do not address the optimization of the state itself, since this has been investigated in detail in previous studies \cite{Mogyorosi2019,Hanamura2025}, where state fidelity or other relevant quantities such as nonlinear squeezing were used as figures of merit. It was observed in Ref. \cite{Hanamura2025} that if $s_0$ is sufficiently large, then the number of detected photons $n$ can be reduced while simultaneously changing the control parameters $s_0$ and $\delta_0$ to produce a  good approximation of the originally targeted state. Reduction of $n$ is desirable, since it typically leads to higher preparation probability. Therefore, when optimizing the state parameters $s_0$, $\delta_0$, and $n$, it appears sensible to seek the lowest $n$ for which the conditionally generated state (\ref{psin})  meets the required performance such as state fidelity or nonlinear squeezing.

\begin{acknowledgments}
This work was supported by Palacký University under Projects No. IGA-PrF-2025-010 and IGA-PrF-2026-005.
\end{acknowledgments}

\section*{Data availability}
The data that support the findings of this article are openly available \cite{Zenodo}.

\end{document}